\documentclass[12pt]{article}

\usepackage{amsfonts,amsmath,amssymb,epsf}
\usepackage{mathtools} 
\usepackage{mathrsfs}
\usepackage{graphics} 
\usepackage{ifpdf}
\ifpdf
\usepackage{epstopdf}
\usepackage[colorlinks=true,
            citecolor=blue,
            linkcolor=blue]
           {hyperref}
\else
\usepackage[hypertex]{hyperref}
\fi
\usepackage{xypic}

\usepackage{graphicx}
\usepackage{amssymb}
\usepackage{hyperref}
\usepackage{euscript}
\usepackage{hhline}
\usepackage[all,arrow,matrix,pdf]{xy}
\usepackage{epstopdf}

\newlength{\fighskip} \fighskip=2pt
\newlength{\figvskip} \figvskip=3pt

 \usepackage{sectsty}
 \sectionfont{\large}

\newcommand\EC   {\EuScript{C}}
\newcommand\ED  {\EuScript{D}}

\newcommand\EF  {\EuScript{F}}
\newcommand\EM  {\EuScript{M}}

\newcommand\Hc  {\mathcal{H}}

\newcommand\vect  {\EuScript{H}\mathrm{ilb}}
\newcommand{\id}      {\mathrm{id}}
\newcommand{\ev}     {\mathrm{ev}}

\usepackage{csquotes}

\begin{document}

\begin{center} \LARGE
Electric-magnetic duality of lattice systems with topological order
\end{center}

\vskip 2em
\begin{center}
Oliver Buerschaper$^{a,b}$,\, Matthias Christandl$^{c}$,\, Liang Kong$^{d,e}$,\, Miguel Aguado$^{b}$,\, 
\\[1em]
\it$^a$ Perimeter Institute for Theoretical Physics, 
31 Caroline Street North, Waterloo, Ontario, Canada N2L 2Y5
\\
$^b$ Max-Planck-Institut f\"ur Quantenoptik,  
Hans-Kopfermann-Stra\ss{}e 1, D-85748 Garching, Germany
\\
$^c$ Institute for Theoretical Physics, ETH Zurich, 8093
  Zurich, Switzerland
\\
$^d$ Institute for Advanced Study (Science Hall) 
Tsinghua University, Beijing 100084, China
\\
$^e$ Department of Mathematics and Statistics
University of New Hampshire, Durham, NH 03824, USA
\end{center}

\vskip 4em
\begin{abstract}
  We investigate the duality structure of quantum lattice systems with
  topological order, a collective order also appearing in fractional
  quantum Hall systems.  We define electromagnetic (EM) duality for
  all of Kitaev's quantum double models based on discrete gauge
  theories with Abelian and non-Abelian groups, and identify its
  natural habitat as a new class of topological models based on Hopf
  algebras.  We interpret these as extended string-net models,
  whereupon Levin and Wen's string-nets, which describe all
  intrinsic topological orders on the lattice with parity and time-reversal
  invariance, arise as magnetic and electric projections of the
  extended models.  We conjecture that all string-net models can be
  extended in an analogous way, using more general algebraic and tensor-categorical
  structures, such that EM duality continues to hold.  We also identify 
  this EM duality with an invertible domain wall. Physical
  applications include topology measurements in the form of pairs of
  dual tensor networks.
\end{abstract}

\newpage 

%%%%%%%%%%%%%%%%%%%%%%%%%%%%%%%%%%%%%%%%%%%%%%%%%%%%%%%%%
\section{Introduction}
\label{sec:intro}

\emph{Duality} ranks among the deepest ideas in physics.  Dualities
are powerful probes into the internal structure of mathematical and
physical theories, including quantum many-body systems and field
theories~\cite{Savit}.  Frequently a duality relates weakly and
strongly coupled regimes of physical systems, providing precious
information beyond the reach of perturbative methods.  Many physical
instances of duality involve the exchange of electric and magnetic
degrees of freedom.  Such a symmetry for the Maxwell equations
\emph{in vacuo} led Dirac to the introduction of pointlike sources of
magnetic field to extend this \emph{electric-magnetic} (EM) duality to
matter, providing a unique argument for the quantisation of electric
charge~\cite{Dirac}.  Dirac's insight has had profound influence in
field theory and beyond (see, e.g.,~\cite{Seiberg}).

\emph{Topological order}~\cite{WenOrder} appears in a fascinating
class of condensed matter phases, notably the fractional quantum Hall
effect.  Topological phases are sensitive only to global properties of
the underlying space: their effective quantum field theories have only
non-local degrees of freedom~\cite{TQFT}.  This non-local order has
deep connections to high-energy physics and a rich mathematical
structure.  In $2D$, it is associated with localised objects with
braiding statistics, that is to say, \emph{anyons}~\cite{WilczekOne,
  WilczekTwo}.  In topological lattice models, these anyons appear as
excitations.

Kitaev's proposal to use topologically ordered quantum lattice systems
to store and process quantum information linked topological order and
quantum computation~\cite{Kitaev}.  His pioneering \emph{quantum
  double} $\mathrm{D} (G)$ models are built in analogy to gauge
theories with finite gauge groups.  In the gauge context, the
topological degrees of freedom are Wilson loops.  Violations of gauge
invariance and magnetic fluxes are described analogously, hinting at
EM duality.  Although there are deeper structures at play, $\mathrm{D}
(G)$ models can be conveniently studied using group theory; there is
an explicit \emph{local} understanding of anyonic excitations and
their interplay in terms of group representations.  Particularly
simple are the models based on Abelian groups, including the toric
code, the first laboratory for many ideas in topological order.  These
models provide us with the first emergence of duality in topologically
ordered systems, namely a self-duality of the toric code and Abelian
$\mathrm{D} (G)$ models that involves the exchange of the direct and
dual lattices; an EM duality, in gauge theory language.  Its
importance has been stressed by Fendley~\cite{Fendley}.

So far the extension of the EM duality of the toric code to
non-Abelian models has remained an open problem, and thus a deep
source of insight into the structure of topological order has been
only available in the very restricted Abelian setting.  The natural
place to look for a well defined EM duality are Kitaev's quantum
double models, since the excitations there are clearly understood and
can be assigned electric and magnetic quantum numbers.

In Sections~\ref{sec:toric}--\ref{sec:em}, we uncover the EM duality for all of Kitaev's $\mathrm{D} (G)$
models, both Abelian and non-Abelian, and beyond.  We do this by
identifying the natural setting of the EM duality as the quantum double
models based on Hopf algebras, a class of models anticipated
in~\cite{Kitaev} and defined and studied in~\cite{BMCA}.  More precisely, we will show that the EM duality identifies the $[\mathrm{D}(H)]_\Lambda$ model, a lattice model built from the structure of a $C^\ast$-Hopf algebra $H$ on a lattice $\Lambda$, with the $[\mathrm{D}(H^\ast)]_{\Lambda^\ast}$ model, where $H^\ast$ is the dual Hopf algebra (see Section~\ref{sec:hopf}) and $\Lambda^\ast$ the dual lattice.

Levin and Wen defined their landmark string-net (SN) models~\cite{LevinWen} from the physical intuition that topological phases correspond to renormalisation group fixed points. An SN model is a lattice model built from the structure of a unitary tensor category $\EC$. Such SN models describe \emph{all} intrinsic, non-chiral topological orders on the lattice with parity and time-reversal invariance. They are, hence, broader in scope than Kitaev's quantum doubles. Similar to quantum double models, excitations of SN models can be described as the superselection sectors of certain local operator algebras~\cite{KK,kong}. A few problems, however, put the string-net models at a disadvantage. First, the introduction of auxiliary spaces is necessary in general to define an excitation in the bulk (or at a boundary/domain wall). Secondly, except for a few cases (e.g. $\EC = \mathrm{Rep}_G$, the category of representations for a finite Abelian group~$G$) the local Hilbert spaces
are not rich enough to support a well-defined splitting of electric and magnetic degrees of freedom. 
Therefore, it is impossible to define a notion of EM duality
for these models in general. These problems can be fixed by considering an extended version of SN models.

In Section~\ref{sec:qdsn}, following \cite{BA}, we Fourier transform a $[\mathrm{D}(H)]_\Lambda$ model to an \emph{extended} string-net (ESN) model, a variant of the SN model (based on $\EC=\mathrm{Rep}_H$) with a canonically enlarged Hilbert space. This Fourier transform is possible due to a powerful mathematical theory called \emph{Tannaka-Krein duality}~\cite{Ulbrich}. Canonically enlarging the Hilbert space of the SN model amounts to supplying the category $\EC$ with an additional structure which is called a fiber functor $\omega\colon \EC \to \vect$, i.e. a monoidal functor from $\EC$ to the category $\vect$ of finite dimensional Hilbert spaces over the complex numbers~$\mathbb{C}$. This fiber functor implies integral quantum dimensions for all simple objects of~$\EC$. We show that the EM dual of an ESN model based on $(\EC, \omega)$ is the ESN model constructed from the dual tensor category $\EC^\ast$ equipped with the dual fiber functor $\omega^\ast$. 

\begin{figure}
\begin{center}
  \includegraphics[scale=1.3]{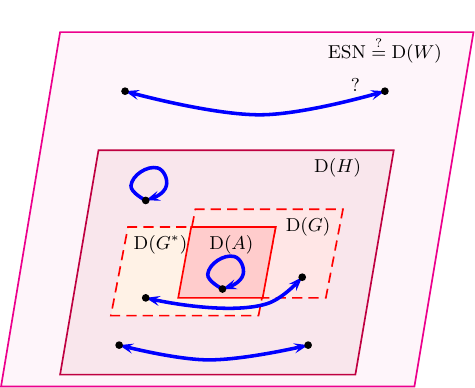}
\end{center}
\caption[Bla]{\textsl{The duality landscape of quantum double models}.  We
  represent: Quantum doubles $\mathrm{D} (A)$ based on Abelian groups,
  including the toric code, all of which are self-dual; general group
  quantum double models $\mathrm{D} (G)$, related by duality to
  quantum double models $\mathrm{D} (G^\ast)$ based on algebras of
  \emph{functions} on groups, whose intersection with $\mathrm{D} (G)$
  is $\mathrm{D} (A)$; these are all instances of quantum double
  models $\mathrm{D} (H)$ based on $C^\ast$-Hopf algebras, a larger
  class closed under duality, with cases of self-duality beyond
  groups~\cite{BMCA}.  We conjecture that EM duality can be defined
  for all extended string-net models $\mathrm{ESN}$, and that these
  can be identified as quantum double models $\mathrm{D} (W)$ based on
  weak $C^\ast$-Hopf algebras.  Arrows denote model identifications by
  EM duality.}
\label{fig:dualities} 
\end{figure}

In Section~\ref{sec:esn-to-sn}, by projecting out the canonically enlarged Hilbert space as in~\cite{BA}, we obtain an SN model, called a magnetic projection, from an ESN model. Importantly, this projection preserves the ground state space. Applying the EM duality, each ESN model maps to two SN models, its electric and magnetic projections (as depicted in Figure~\ref{fig:schememodels}). The EM duality at the level of ESN models induces a duality between these two SN models. Hence, these two SN models describe the same topological phase. Yet, magnetic projections of the above ESN models do not cover all SN models (because fiber functors do not exist for categories~$\EC$ beyond $\mathrm{Rep}_H$).
We conjecture that (i)~\emph{all} SN models are still obtained by magnetic projection from parent ESN models which are based on \emph{weak} $C^\ast$-Hopf algebras, and (ii)~EM duality continues to hold in these generalized ESN models. These results lead us to propose the duality landscape of Figure~\ref{fig:dualities}.

At the level of SN models, dualities between two models are equivalences between two systems of bulk excitations. Mathematically, they correspond to braided equivalences of unitary modular tensor categories. As shown in \cite{KK}, they can be further identified with invertible (or transparent) domain walls. We will show in Section~\ref{sec:defect} that the EM duality between the electric and magnetic projections of an ESN model is canonically associated to a certain invertible domain wall. The corresponding domain wall at the level of ESN models is also briefly discussed.

In Section~\ref{sec:phys}, we discuss a physical application of the EM duality in topology measurements in the form of pairs of dual tensor networks.

%%%%%%%%%%%%%%%%%%%%%%%%%%%%%%%%%%%%%%%%%%%%%%%%%%%%%%%%%
\section{The toric code and self-duality}
\label{sec:toric} 

Let us begin by reviewing the well-known self-duality in the toric
code~\cite{Kitaev}, the quantum double model based on the group
$\mathbb{Z}_2$.  This is a model of qubits along the edges of a $2D$
lattice $\Lambda$, with a Hamiltonian of the form
\begin{equation}\label{generic_hamiltonian}
  \mathcal{H}
=
  - \, \sum_v A_v - \, \sum_p B_p
\; ,
\end{equation}
where $v$ runs over vertices and $p$ runs over plaquettes of
$\Lambda$.  Mutually commuting vertex and plaquette operators are
projectors involving Pauli matrices:
\begin{equation}\label{toric_code_projectors}
  A_v
=
  \frac{
    1 + \otimes_{ i \in v } \sigma^x_i
  }{
    2
  }
\; , \quad
B_p
=
  \frac{
    1 + \otimes_{ j \in p } \sigma^z_j
  }{
    2
  }
\; ,
\end{equation}
with support on the edges around the corresponding vertex or
plaquette.  Ground states of the frustration-free Hamiltonian
\eqref{generic_hamiltonian} minimise each term in the sum, that is,
they satisfy all vertex and plaquette constraints $A_v | \psi \rangle
= | \psi \rangle$, $B_p | \psi \rangle = | \psi \rangle$.  Breakdown
of any one such constraint means that the state is in an excited
level, and can be interpreted as the presence of a localised
excitation, or `particle', sitting at the vertex or the plaquette
whose constraint is broken; from the gauge theory interpretation,
these are called electric and magnetic excitations, respectively.

Self-duality means that the global unitary
\begin{equation}\label{}
  U_\Lambda
=
  \otimes_{ i \in \Lambda } U_i
\; , \quad
  U_i
=
  \frac{ ( \sigma^x + \sigma^z )_i }{ \sqrt{2} }
\; ,
\end{equation}
built from Hadamard maps on each edge $i$, maps the toric code on
$\Lambda$ to a toric code on the dual lattice $\Lambda^\ast$:
\begin{equation}\label{hadamard_for_toric_code}
  U_\Lambda \, A_v \, U_\Lambda^\dagger
=
  \widetilde{B}_v
\; , \quad
  U_\Lambda \, B_p \, U_\Lambda^\dagger
=
  \widetilde{A}_p
\; ,
\end{equation}
where now $\widetilde{B}_v$ corresponds to dual plaquette $v$, and
$\widetilde{A}_p$ to dual vertex $p$ in $\Lambda^\ast$.  The
electric-magnetic nature of duality~\eqref{hadamard_for_toric_code}
comes from the interchange of vertices and plaquettes in going from
the original to the dual lattice.

Hidden in the action of the global Hadamard map $U_\Lambda$ are $(a)$
the mapping from $\Lambda$ to $\Lambda^\ast$, with the corresponding
reinterpretation of vertex and plaquette operators, and $(b)$ a
mapping of the elements of the group $\mathbb{Z}_2$ to the functions
from this group to the scalars.  We obtain again a toric code on the
dual lattice because, for an \emph{Abelian} group, the space of
functions has the same structure as the group algebra, which is the
space of linear combinations of group elements.  One can exploit this
algebraic fact to extend the self-duality of the toric code to all
quantum double models based on Abelian groups in a straightforward
way: this is the $\mathrm{D} (A)$ region of
Figure~\ref{fig:dualities}.

%%%%%%%%%%%%%%%%%%%%%%%%%%%%%%%%%%%%%%%%%%%%%%%%%%%%%%%%%
\section{Quantum double models based on groups}
\label{sec:generaldg}

\begin{figure}
\begin{center}
  \includegraphics[scale=1.3]{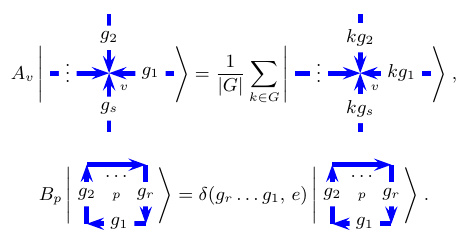}
\end{center}
\caption{\textsl{Projectors in the $\mathrm{D} (G)$ model.}  Vertex
  ($A_v$) and plaquette ($B_p$) projectors in the Hamiltonian of
  Kitaev's $\mathrm{D} (G)$ model are defined by their action on a
  basis of the corresponding multi-edge Hilbert space.  Oriented edges
  are labelled by group elements, denoting kets in the orthonormal
  computational basis $\{ | g \rangle \}_{ g \in G }$; change of
  orientation is implemented by applying the inversion map $g \mapsto
  g^{-1}$.  The vertex operator projects onto gauge-invariant
  configurations, that is, symmetrises over simultaneous
  multiplication with group elements.  The plaquette operator involves
  a Kronecker delta on the unit element $e \in G$ and projects onto
  trivial magnetic flux across $p$.}
\label{fig:vertexplaquettedg} 
\end{figure}

Let us now define general quantum double models $\mathrm{D} (G)$ based
on groups.  Here, we start from a lattice with oriented edges, whose
Hilbert spaces have an orthonormal basis $\{ | g \rangle \}_{ g
  \in G }$ labelled by elements of a finite group $G$.  The
Hamiltonian of the $\mathrm{D} (G)$ model is still of the
form~\eqref{generic_hamiltonian}, but now the mutually commuting
vertex and plaquette operators are the projectors defined in
Figure~\ref{fig:vertexplaquettedg}.
From the gauge theory point of view, edge degrees of freedom
correspond to parallel transport operators taking values in the gauge
group, vertex operators project onto gauge invariant configurations,
and plaquette projectors minimise the Wilson action, i.~e., they
project onto configurations with trivial magnetic flux across the
plaquette.

Excitations of the $\mathrm{D} (G)$ models are classified
algebraically.  Magnetic excitations, sitting on plaquettes,
correspond to conjugacy classes of the gauge group; electric charges,
sitting on vertices, are labelled by irreducible representations of
$G$; and there are excitations, called dyons, living at the
combination of a plaquette and a vertex and having both a magnetic
part (a conjugacy class $C$ of $G$) and an electric part (an
irreducible representation of the centraliser of $C$, which is a
subgroup of $G$).  In terms of more general algebraic structures, all
these objects are just the irreducible representations of a
quasitriangular Hopf algebra, Drinfel'd's \emph{quantum double}
$\mathrm{D} (G)$ \cite{drinfeld}, and the distribution of topological
charges over the lattice can be characterised by its reaction to
operators representing this algebra, of which the vertex and plaquette
projectors are examples.

The obvious parallelism of plaquettes and vertices in the Abelian case
is lost if $G$ is non-Abelian.  Upon switching to the dual lattice, we
cannot reinterpret the operators $A_v$ and $B_p$ as, respectively,
plaquette and vertex operators in a quantum double model \emph{based
  on a group}, since the functions on a non-Abelian group no longer
have the same structure as the group algebra itself.  Hence, the
extension of the EM duality to non-Abelian $\mathrm{D} (G)$ models is
not possible within the class of quantum doubles based on groups.  Yet
dual models \emph{can} be constructed which are quantum doubles
based no longer on groups, but rather on algebras of functions; this
is shown as the arrow between the regions $\mathrm{D} (G)$ and
$\mathrm{D} (G^\ast)$ in Figure~\ref{fig:dualities}.

To make sense of this, one has to widen the construction of quantum
double lattice models beyond the group case.  As we will show, the
natural habitat for the EM duality of $\mathrm{D} (G)$ models is the
class of quantum double models based on \emph{Hopf
  algebras}~\cite{BMCA}; this is the region~$\mathrm{D} (H)$ in
Figure~\ref{fig:dualities}.  This is the smallest class containing all
the $\mathrm{D} (G)$ models that is closed under tensor products and
EM duality.  The duality, moreover, takes a remarkably simple form in
the language of Hopf algebras.

%%%%%%%%%%%%%%%%%%%%%%%%%%%%%%%%%%%%%%%%%%%%%%%%%%%%%%%%%

\section{The class of Hopf algebras}
\label{sec:hopf} 

When studying quantum many-body systems, more general notions of
symmetry emerge than those furnished by group theory.  In particular,
linear transformations acting on tensor products of vector spaces lead
naturally to Hopf algebras.  For a detailed account of these in our
context we refer to~\cite{BMCA}.  In the following we give an
intuitive grasp of their structure, which is necessary to understand
the $\mathrm{D} (H)$ lattice models.

We regard a Hopf algebra $H$ as a space of transformations on a
many-body Hilbert space.  First of all, the linear nature of the
target is naturally extended to its transformations, so Hopf algebras
are vector spaces.  We must be able to compose transformations and to
include the identity transformation, so $H$ has a multiplication of
vectors, and a unit, making it into an algebra.  Most importantly, we
must have a rule to \emph{distribute} the action of an element of $H$
into a tensor product of target spaces; this is the so-called
comultiplication.  Additionally $H$ has a trivial representation
$\varepsilon$, called counit, making precise the notion of spaces
invariant under the action of $H$.  In the same way as for groups, the
representation theory of Hopf algebras includes a notion of conjugate
representation, implemented via an antipode mapping, which for groups
is just the inversion $g \mapsto g^{-1}$.

To be able to construct Hilbert spaces we use finite-dimensional Hopf
$C^\ast$-algebras, where an inner product can be defined.  In
addition, they come equipped with a canonical, normalised, highly
symmetric element, the \emph{Haar integral} $h \in H$, invariant under
multiplication in the sense $a \cdot h = \varepsilon(a) \, h$ for all
elements $a$ of $H$.  This canonical element is crucial for the
construction of the lattice model and its ground states.

\begin{figure}
\begin{center}
  \includegraphics[scale=1.3]{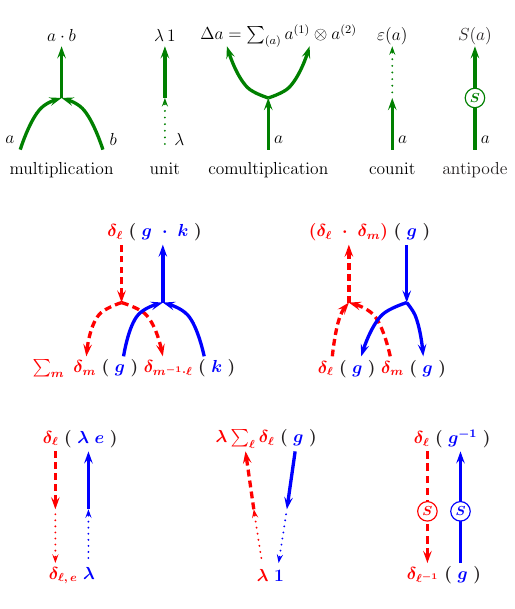}
\end{center}
\caption{\textsl{Structure and dualities in Hopf algebras.}  The
  \emph{first row} represents the structure maps of a Hopf algebra
  $H$.  The \emph{second and third rows} illustrate the algebraic
  duality within the class of Hopf algebras, taking as examples a
  group algebra $H = \mathbb{C} G$ and its dual $H^\ast$, whose
  structure is defined implicitly in the figure.}
\label{fig:hopf} 
\end{figure}

A graphical language which is particularly advantageous for the
treatment of Hopf algebras is outlined with reference to the first row
of Figure~\ref{fig:hopf}. Each diagram in the first row represents one
of the structure maps of a Hopf algebra as introduced above:
multiplication, unit, comultiplication, counit, and antipode, all
viewed as linear mappings from a vector space at the bottom of the
diagram to a vector space at the top of the diagram. Tensor factors of
these vector spaces are represented by juxtaposition.

The root of the EM duality to be unveiled in the following is an
algebraic fact: \emph{The class of finite-dimensional Hopf
  $C^\ast$-algebras is closed under dualisation}.  That is, given a
finite-dimensional $C^\ast$-Hopf algebra $H$, its dual space $H^\ast$
(the \emph{functions} from $H$ to the scalars) is again a
finite-dimensional $C^\ast$-Hopf algebra, whose structure is,
moreover, determined by the structure of $H$.  This closure property
is shared by the class of Abelian group algebras, which are all
self-dual, but not by the whole class of group algebras.  The
landscape of Figure~\ref{fig:dualities} reflects these statements in
the world of lattice models.

With reference to the second and third rows of Figure~\ref{fig:hopf},
we use the case where the Hopf algebra is a group algebra, $H =
\mathbb{C} G$ to illustrate the algebraic duality in the class of Hopf
algebras. For each diagram, the upper expression is equal to the lower
expression; the structure maps of $H$ and $H^\ast$ are represented by
solid and dashed lines, respectively, according to the conventions of
the first row.  Hence, the comultiplication in $H^\ast$ is determined
by the multiplication in $H$, the multiplication in $H^\ast$ is
determined by the comultiplication in $H$, the counit in $H^\ast$ is
determined by the unit in $H$, the unit in $H^\ast$ is determined by
the counit in $H$, and the antipode in $H^\ast$ is determined by the
antipode in $H$; this duality holds for general finite-dimensional
Hopf algebras.  We use the form of the structure maps in the basis $\{
g; \, g \in G \}$ of the group algebra, and the basis $\{ \delta_\ell;
\, \ell \in G \}$ of its dual, where $\delta_\ell \colon g \mapsto \delta_{
  \ell, \, g }$ are Kronecker delta functions.  As for the extra
structure needed to define our lattice models, the $C^\ast$-algebra
structures of $H$ and $H^\ast$ are defined by declaring $\{ g \}$ to
be an orthonormal basis, and $\{ \delta_\ell \}$ to be orthonormal up
to a factor, $( \delta_\ell, \, \delta_m ) = | G |^{-1} \, \delta_{
  \ell, \, m }$.  The Haar integrals $h$ and $\phi$ in the group
algebra and its dual are defined in equation~\eqref{haargroups} below.

%%%%%%%%%%%%%%%%%%%%%%%%%%%%%%%%%%%%%%%%%%%%%%%%%%%%%%%%%
\section{The $\mathrm{D} (H)$ lattice models}
\label{sec:dhmodels} 

\begin{figure}
\begin{center}
  \includegraphics[scale=1.3]{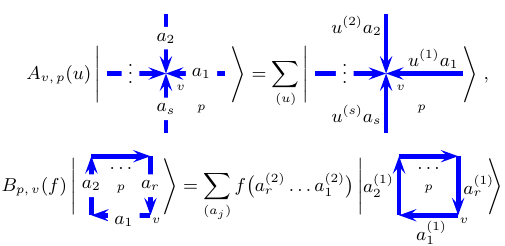}
\end{center}
\caption{\textsl{The core of $\mathrm{D} (H)$ models}.  Topological
  models based on Hopf algebras are defined via these vertex and
  plaquette representations of elements $u$ of Hopf algebra $H$ and
  elements $f$ of the dual $H^\ast$.}
\label{fig:vertexplaquettedh} 
\end{figure}

Quantum double models based on Hopf algebras, $\mathrm{D} (H)$ models
for short, are a class of topological models, defined on $2D$ lattices
whose local Hilbert space along oriented edges is a finite-dimensional
$C^\ast$-Hopf algebra $H$.  Figure~\ref{fig:vertexplaquettedh} defines
a set of operators on this lattice which depend on elements $u \in H$
and $f \in H^\ast$.  Oriented edges carry, in general, elements of the
finite-dimensional $C^\ast$-Hopf algebra $H$.  Orientation can be
reversed by using the antipode map $S$.  The sums in the formulas of
Figure~\ref{fig:vertexplaquettedh} denote the appropriate
\emph{comultiplication} or splitting of algebra elements $u$ into
several tensor factors $\sum u^{(1)} \otimes u^{(2)} \otimes \ldots
\otimes u^{(s)}$.  Vertex operators $A_{v, \, p}$ now depend also on a
plaquette $p$ marking the beginning of a loop around the vertex $v$,
and correspondingly for the plaquettes.  These operators probe the
algebraic structure of the models: Given overlapping vertex $v$ and
plaquette $p$, the corresponding vertex operator $A_{v, \, p} (u)$ and
the plaquette operator $B_{p, \, v} (f)$ interact nontrivially with
each other, building representations of Drinfel'd's quantum double
algebra $\mathrm{D} (H)$~\cite{drinfeld}.  The irreducible
representations of $\mathrm{D} (H)$ classify the superselection
sectors, and thus the anyonic excitations, of the quantum double
lattice model~\cite{Kitaev}\footnote{Technically, the $A_{ v, \, p } (
  u )$ are representations of $H$, and the $B_{ p, \, v } ( f )$ are
  representations of a variant of the dual; namely, the dual $( H^{
    \mathrm{op} } )^\ast$ of the \emph{opposite} algebra of $H$ (where
  the multiplication is flipped)}.

When elements $u$ and $f$ in Figure~\ref{fig:vertexplaquettedh} are
taken as the canonical Haar integrals $h \in H$ and $\phi \in H^\ast$,
respectively, the resulting operators
\begin{equation}\label{ophamiltdh}
  A_v = A_{ v, \, p } ( h )
\; , \quad
  B_p = B_{ p, \, v } ( \phi )
\end{equation}
are mutually commuting projectors defining
via~\eqref{generic_hamiltonian} the Hamiltonian of the $\mathrm{D}
(H)$ model~\cite{BMCA}.  The Hamiltonian is thus entirely constructed
from the canonical elements $h$, $\phi$ and the algebraic structure of
$H$.

Kitaev's models constitute a subclass of $\mathrm{D} (H)$ models,
recovered when $H$ is the group algebra of a finite group $G$.  Both
this algebra $\mathbb{C} G$ and its dual $G^\ast$ have particularly
simple $C^\ast$-Hopf algebra structures, summarised in the second and
third rows of Figure~\ref{fig:hopf}.  In particular, their Haar
integrals read
\begin{equation}\label{haargroups}
  h
=
  \frac{ 1 }{ | G | } \,
  \sum_{ g \in G } g
\in
  \mathbb{C} G
\; , \quad
  \phi = \delta_e \in G^\ast
\; .
\end{equation}
Taking into account this equation, operators in
Figure~\ref{fig:vertexplaquettedg}, defining the Hamiltonian of the
$\mathrm{D} (G)$ model, can be checked to coincide with the general
expression~\eqref{ophamiltdh}.

%%%%%%%%%%%%%%%%%%%%%%%%%%%%%%%%%%%%%%%%%%%%%%%%%%%%%%%%%
\section{Electric-magnetic duality}
\label{sec:em}

\begin{figure}
\begin{center}
  \includegraphics[scale=1.3]{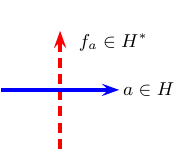}
\end{center}
\caption{\textsl{Going over to the dual lattice}. The mapping $U\colon H
  \rightarrow H^\ast$ is here associated with a lattice dualisation
  $\Lambda \rightarrow \Lambda^\ast$.}
\label{fig:dualbondvar} 
\end{figure}

We now define EM duality for general $\mathrm{D} (H)$ models.
Consider the following unitary map $U$ from $H$ to its dual $H^\ast$:
\begin{equation}\label{emduality}
  a \stackrel{U}{\longmapsto} f_a \; ,
\quad
  f_a (b)
=
  \sqrt{ \dim H  } \,
  \phi ( a \cdot b )
\; ,
\end{equation}
where the function $f_a$ is constructed via the dual Haar integral $\phi \in H^\ast$ (for
instance, for groups, $f_g = \sqrt{ | G | } \, \delta_{ g^{-1} }$).
We associate this mapping with the transformation of $\Lambda$ into
its dual lattice $\Lambda^\ast$ as shown in
Figure~\ref{fig:dualbondvar}.  In this figure, the solid edge is in
$\Lambda$, while the dashed edge is its dual in $\Lambda^\ast$.  The
orientation of the dual edge is chosen by convention so that it points
to the left of the direct edge.  The degree of freedom $a \in H$ in
the Hilbert space $H$ of the edge in $\Lambda$ becomes a degree of
freedom $f_a$ of the Hilbert space $H^\ast$ of the dual edge.  When
applied simultaneously on all edges of $\Lambda$, this mapping
implements EM duality.

The vertex and plaquette representations in
Figure~\ref{fig:vertexplaquettedh}, which in particular determine the
Hamiltonian of the $\mathrm{D} (H)$ model on $\Lambda$, are mapped by
the global operation $U_\Lambda = \bigotimes_{ i \in \Lambda } U_i$
into precisely the representations associated with the $\mathrm{D} (
H^\ast )$ model on the dual lattice
$\Lambda^\ast$\footnote{Technically, representations $A_{ v, \, p } (
  u )$ of $H$ and $B_{ p, \, v } ( f )$ of $( H^{ \mathrm{op} }
  )^\ast$ act on edges of $\Lambda$, that is, spaces of the form $H^{
    \otimes n }$.  They get mapped by $U_\Lambda$ to representations
  $\widetilde{B}_{ v, \, p } ( u )$ of $( ( H^\ast )^{ \mathrm{op} }
  )^\ast$ and $\widetilde{A}_{ p, \, v } ( f )$ of $H^\ast$ on edges
  of $\Lambda^\ast$, that is, spaces of the form $( H^\ast )^{ \otimes
    n}$.  The latter, for coinciding $v$ and $p$, define
  representations of the $\mathrm{D} ( H^\ast )$ algebra, building the
  $\mathrm{D} ( H^\ast )$ model on $\Lambda^\ast$.}:
\begin{equation}\label{maprepresentationstodual}
  U_\Lambda \,
  B_{ p, \, v } ( f ) \,
  U_\Lambda^\dagger
=
  { \widetilde A }_{ p, \, v } ( f )
\; , \quad
  U_\Lambda \,
  A_{ v, \, p } ( u ) \,
  U_\Lambda^\dagger
=
  { \widetilde B }_{ v, \, p } ( u )
\; ,
\end{equation}
where $u \in H$ and $f \in H^\ast$.  Here $p$ is a plaquette of
$\Lambda$ and a vertex of $\Lambda^\ast$; $v$ is a vertex of $\Lambda$
and a plaquette of $\Lambda^\ast$.
Expression~\eqref{maprepresentationstodual} generalises
equation~\eqref{hadamard_for_toric_code}, singling out $U_\Lambda$ as
the EM duality mapping.

Thus, $U_\Lambda$ \emph{identifies} the $\mathrm{D} (H)$ model on
$\Lambda$ with the $\mathrm{D} ( H^\ast )$ model on $\Lambda^\ast$, as
it unitarily transforms the Hilbert spaces and Hamiltonian of the former into
those of the latter.  We write this symbolically as $[ \mathrm{D} (H)
]_\Lambda = [ \mathrm{D} ( H^\ast ) ]_{ \Lambda^\ast }$.

But EM duality is deeper than a mere identification of energy levels.
The Hamiltonian involves just the operators $A_v ( h )$ and $B_p (
\phi )$, but, according to~\eqref{maprepresentationstodual},
$U_\Lambda$ induces a transformation of all operators of
Figure~\ref{fig:vertexplaquettedh} that respects their algebra
structure, and relates the anyonic excitations of both models.

%%%%%%%%%%%%%%%%%%%%%%%%%%%%%%%%%%%%%%%%%%%%%%%%%%%%%%%%%
\section{From quantum doubles to extended string-nets}
\label{sec:qdsn} 
  
In~\cite{BA} the $\mathrm{D} (G)$ model was identified, via a Fourier transformation, with an \emph{extended string-net model} on the same lattice. In particular, the edge degrees of freedom can be spanned by a Fourier basis 
\begin{equation}
|\mu a b\rangle = \sqrt{\frac{\dim \mu}{|G|} }  \sum_{g\in G}\,\, D^\mu(g)_{ab} |g\rangle,
\end{equation}
where $\mu$ runs over the irreducible representations (irreps) of $G$ and $D^\mu$ is a fixed matrix realization of the irreducible representation $\mu$. Using this new basis, $A_v$ and $B_p$ operators in the Hamiltonian can both be expressed in terms of certain projectors \cite[Eq.(11)(17)]{BA} as follows: 
\begin{equation} \label{eq:A-v-ba}
A_v |\{ \mu_ia_ib_i\}\rangle_v = \sum_{c_1, \cdots, c_r} W_{\{c_i\}, \{b_i\}}^{ \{ \mu_i \}} |\{ \mu_ia_ib_i\}\rangle_v
\end{equation}
where 
\begin{equation*}
W_{\{c_i\}, \{b_i\}}^{ \{ \mu_i \}} = \frac{1}{|G|} \sum_{l\in G} \prod_i D^{\mu_i}(l)_{c_ib_i}
\end{equation*}
is the projector onto the trivial isotopic subspace of $\otimes_i \mu_i$, and 
\begin{equation} \label{eq:B-p-ba}
B_p \coloneqq \sum_\nu \frac{ \dim \nu}{|G|} B_p(\nu)
\end{equation}
with $B_p(\nu)$ acting on a hexagon-shaped plaquette as follows: 
\begin{equation}  \label{eq:B-p-nu-ba}
\langle \{ \mu_j'a_j'b_j'\} | B_p(\nu) | \{ \mu_ja_jb_j\} \rangle = \prod_{i=1}^6 \sqrt{\dim \mu_i \dim \mu_i'} \sum_{c_1, \cdots, c_6} \prod_{i=1}^6 W_{a_ia_i'c_i, b_ib_i'c_{i+1}}^{\mu_i\mu_i'\nu}
\end{equation}
where the index $i$ is cyclic, i.e. $i+1=1$ when $i=6$.

The representation theory of finite-dimensional $C^\ast$-Hopf algebras is essentially identical to that of finite groups, the only real difference being that the fusion of representations need not be commutative. Therefore the construction of~\cite{BA} generalizes to any $\mathrm{D} (H)$ model.  That is, if $H$ has matrix
irreps $D^\mu$ associated to irrep $\mu$, we define a Fourier basis in $H$ containing the following algebra elements:
\begin{equation}\label{fourierbasish}
  | \mu a b \rangle
=
  \sqrt{ \frac{ \dim \mu }{\dim H } } \,
  \sum_{ (h) }
  D^\mu ( h^{(1)} )_{ a b } \,
  h^{(2)}
\; ,
\end{equation}
where $a, \, b = 1, \, \ldots, \, \dim \mu $ are matrix indices for irrep $D^\mu$.  This is an orthonormal basis for the Hilbert spaces at each edge. The analysis of~\cite{BA} carries through intact to show that the $[ \mathrm{D} (H) ]_\Lambda$ model can be written as an extended string-net model $[ \mathrm{ESN}_{(\EC, \omega)} ]_\Lambda$, where the meaning of $(\EC, \omega)$ will be explained later, with edge degrees of freedom labeled by triplets $|\mu a b \rangle$.

Readers might wonder why we give the $[\mathrm{D} (H)]_\Lambda$ model the new name \enquote{extended string-net model $[ \mathrm{ESN}_{(\EC, \omega)} ]_\Lambda$}. The reason is that the above construction for the $[\mathrm{D} (H)]_\Lambda$ model can be reformulated completely in terms of tensor categorical notions, i.e. a pair $(\EC, \omega)$ (explained later), without referring to groups or Hopf algebras at all. This reformulation, which will be given explicitly below, is important and deserves a new name because we believe that the tensor-categorical structures should be viewed as more fundamental than groups or Hopf algebras, and the latter notions should be viewed as symmetries emerging from the former notions. Such a point of view was expressed earlier in \cite{LevinWen} and will be made more precise in this article. Mathematically, this reformulation is possible due to the so-called Tannaka-Krein duality \cite{Ulbrich} which explains the equivalence between the notion of a Hopf algebra and that of a tensor category $\EC$ equipped with a fiber functor $\omega$. 

In the rest of this section we outline the definition of ESN models purely from tensor-categorical structures. The developments in the following sections will be presented in terms of the quantum double and/or the tensor-categorical formulation where appropriate.

We will start from a unitary finite fusion category $\EC$ with simple objects $\mu, \nu, \lambda, \ldots \in I$, where $I$ is a finite set, and a simple unit $1_\EC \in I$. Associating auxiliary matrix indices with each simple object in $\EC$ amounts to assigning a vector space $\omega(\mu)$ to each simple object $\mu\in I$. 
In the quantum double model picture, $\mu$ is the label for an irreducible representation of a finite group or a Hopf algebra, and $\omega(\mu)$ is nothing but the representation space associated with $\mu$. 
It is easy to generalize this assignment to all objects in $\EC$ by requiring $\mu \oplus \nu \mapsto \omega(\mu) \oplus \omega(\nu)$ for $\mu, \nu\in I$
so that we obtain a functor $\omega\colon \EC \to \vect$, where $\vect$ is the category of finite dimensional Hilbert spaces. The assignment $\omega(\mu)$ cannot be arbitrary. The consistency condition needed for the construction of our lattice models is that the functor $\omega\colon \EC \to \vect$ is a monoidal functor. In particular, it means that there are isomorphisms 
\begin{equation}  \label{eq:phi-2}
\phi_2\colon \omega(\mu \otimes \nu) \xrightarrow{\simeq} \omega(\mu) \otimes \omega(\nu)
\end{equation} 
and $\phi_0\colon\omega(1_\EC) \xrightarrow{\simeq} \mathbb{C}$ satisfying certain coherence conditions. Such a functor $\omega$ is called a  fiber functor of $\EC$.

We would like to build a Hamiltonian model on a trivalent lattice with oriented edges. 
Note that the restriction to trivalent lattices (instead of more general planar lattices) is merely a matter of convenience. 
We associate  to each oriented internal edge $e$ a Hilbert space 
\begin{equation} \label{eq:hilbert-e}
\Hc_e= H_{(\EC, \omega)} = \oplus_{\mu\in I} \,\, \omega( \mu ) \otimes \omega ( \mu )^\ast,
\end{equation}
where $\omega(\mu)^\ast$ is the dual vector space of $\omega(\mu)$. 
The total Hilbert space is just $\Hc = \otimes_e \Hc_e$. 

The Hamiltonian is still defined as:
\begin{equation}
H = -\sum_v  A_v   - \sum_p  B_p ,
\end{equation}
where $v$ labels the trivalent vertices and $p$ labels the plaquettes. The operator $A_v$ for each trivalent vertex $v$ (with all edges pointing towards $v$) is a linear map: 
$\otimes_{i=1}^3 \omega(\mu_i)^\ast \otimes \omega(\mu_i)  \rightarrow \otimes_{i=1}^3 \omega(\mu_i)^\ast \otimes \omega(\mu_i)$
defined by: 
\begin{equation}  \label{eq:A-v}
A_v \coloneqq P_{\omega(\mu_1)\omega(\mu_2)\omega(\mu_3)} \otimes \id_{\omega(\mu_1)^\ast} \otimes  \id_{\omega ( \mu_2 )^\ast}  \otimes \id_{\omega ( \mu_3 )^\ast} ,
\end{equation}
where $P_{\omega(\mu_1)\omega(\mu_2)\omega(\mu_3)}\colon \otimes_{i=1}^3 \omega(\mu_i) \rightarrow \otimes_{i=1}^3 \omega(\mu_i)$  is intuitively the projection onto the vacuum channel. However, we do not know if $\EC$ is the category of representations of any group or Hopf algebra a priori. We must therefore define $P_{\omega(\mu_1)\omega(\mu_2)\omega(\mu_3)}$ independently in terms of the data in $\EC$ and the fiber functor $\omega$. Note that the fusion rule in $\EC$ gives a canonical projector morphism $P_{\mu_1\mu_2\mu_3}\colon \mu_1 \otimes \mu_2 \otimes \mu_3 \to \oplus 1_\EC$ in $\EC$ that maps $\mu_1 \otimes \mu_2 \otimes \mu_3$ to the subobject consisting of only a direct sum of copies of the unit object $1_\EC$. Since $\omega$ is monoidal, using \eqref{eq:phi-2}, we obtain a canonical isomorphism 
\begin{equation}    \label{eq:phi-3}
\phi_3\colon \omega(\mu_1)\otimes \omega(\mu_2) \otimes \omega(\mu_3) \xrightarrow{\cong} \omega( \mu_1 \otimes \mu_2 \otimes \mu_3).
\end{equation}
Then we define $P_{\omega(\mu_1)\omega(\mu_2)\omega(\mu_3)} \coloneqq \phi_3^{-1} \circ \omega(P_{\mu_1\mu_2\mu_3}) \circ \phi_3$. It is clear that $P_{\omega(\mu_1)\omega(\mu_2)\omega(\mu_3)}$ so defined is a projector. 

The operator $B_p$ for each plaquette $p$ (taken, for convenience only, as a hexagon) is a map: 
\begin{equation*}
B_p\colon    \otimes_{i=1}^6  ( \oplus_{\lambda_i } \omega(\lambda_i)^\ast \otimes \omega(\lambda_i)  )  \rightarrow
\otimes_{i=1}^6  (\oplus_{\mu_i } \omega(\mu_i)^\ast \otimes \omega(\mu_i) )
\end{equation*}
which can be viewed as an element in the space
\begin{equation}  \label{eq:b-p}
\oplus_{ \{\mu_i\} } \oplus_{ \{\lambda_i\} }  \otimes_{i=1}^6  (  \omega(\mu_i)^\ast \otimes \omega(\mu_i)  \otimes \omega(\lambda_i^\ast)^\ast \otimes \omega(\lambda_i^\ast) )
\end{equation}
where we have used $\omega(\lambda^\ast) \cong \omega(\lambda)^\ast$ because $\omega$ is monoidal. 
We define this element $B_p$ to be 
\begin{equation} \label{eq:B-p}
B_p = \sum_{\nu \in I} \frac{\dim \omega(\nu)}{\dim \Hc_e} \,\, B_p(\nu),
\end{equation}
where $B_p(\nu)$ is again an element in the space of \eqref{eq:b-p}. We define $B_p(\nu)$ in the following way.  First we have the element 
\begin{equation*}
\oplus_{ \{\mu_i\} } \oplus_{ \{\lambda_i\} } \otimes_{i=1}^6 \sqrt{ \dim \mu_i \dim \lambda_i} \,  P_{\omega(\mu_i) \omega(\lambda_i^\ast) \omega(\nu_i) },    
\end{equation*}
where all $\nu_i =\nu$,  in the space
\begin{equation} \label{eq:b-p-precont}
\oplus_{ \{\mu_i\} } \oplus_{ \{\lambda_i\} }  \otimes_{i=1}^6  (  \omega(\mu_i)^\ast \otimes \omega(\mu_i)  \otimes \omega(\lambda_i^\ast)^\ast \otimes \omega(\lambda_i^\ast)  \otimes 
\omega(\nu_i) \otimes \omega(\nu_i)^\ast ).
\end{equation} 
Using the evaluation map $\ev_i\colon \omega(\nu_i)^\ast \otimes \omega(\nu_{i+1}) \rightarrow \mathbb{C}, i=1, \dots, 6$ where $\nu_7=\nu_1$,  we obtain a map $\otimes_{i=1}^6  \ev_i$ from the space \eqref{eq:b-p-precont} to the space \eqref{eq:b-p}.  Then we define
\begin{equation}  \label{eq:B-p-nu}
B_p(\nu) \coloneqq \otimes_{i=1}^6  \ev_i (\oplus_{ \{\mu_i\} } \oplus_{ \{\lambda_i\} } (\otimes_{i=1}^6 \sqrt{ \dim\mu_i  \dim \lambda_i} \, P_{\omega(\mu_i) \omega(\lambda_i^\ast) \omega(\nu_i) } ) ).
\end{equation}

If $\EC=\mathrm{Rep}_H$ for some $C^\ast$-Hopf algebra $H$, then we simply take $\omega$ to be the forgetful functor $F\colon \mathrm{Rep}_H \to \vect$. Then the Hilbert space $H_{(\EC, \omega)}$ (recall Eq. \eqref{eq:hilbert-e}) is isomorphic to the underlying Hilbert space of the $C^\ast$-Hopf algebra $H$. Moreover, it is easy to check that the definition of $A_v$ and $B_p$ (Eq.\,\eqref{eq:A-v}, \eqref{eq:B-p}, \eqref{eq:B-p-nu}) coincides exactly with that of $A_v$ and $B_p$ (see for example Eq. \eqref{eq:A-v-ba}, \eqref{eq:B-p-ba}, \eqref{eq:B-p-nu-ba} in $\mathrm{D}(G)$ cases) in $\mathrm{D}(H)$ models. Therefore, we have 
\begin{equation*}
[\mathrm{D} (H)]_\Lambda = [\mathrm{ESN}_{(\mathrm{Rep}_H, F)}]_\Lambda.
\end{equation*}
Conversely, for any unitary finite fusion category $\EC$ admitting a fiber functor $\omega$, the Hilbert space $H_{(\EC, \omega)}$ is automatically equipped with the structure of a semisimple $C^\ast$-Hopf algebra such that $\EC\simeq \mathrm{Rep}_{H_{(\EC, \omega)}}$. This result is nothing but the statement of the so-called Tannaka-Krein duality \cite{Ulbrich}.

\medskip
The EM-duality for ESN models can be defined in terms of tensor-categorical notions as well. On the one hand, given a pair $(\EC, \omega)$, the category $\vect$ is automatically a $\EC$-module. Then we can define the dual tensor category 
\begin{equation*}
\EC^\ast\coloneqq \mathrm{Fun}_\EC(\vect, \vect)^{\otimes_{\mathrm{op}}}
\end{equation*}
by the category of $\EC$-module functors from $\vect$ to $\vect$ with tensor product defined by the opposite composition of functors. The dual tensor category $\EC^\ast$ is again a unitary finite fusion category. Moreover, $\vect$ is naturally equipped with a $\EC^\ast$-module structure, which further gives rise to a fiber functor \cite{ostrik} 
$\omega^\ast\colon \EC^\ast \to \vect$. More explicitly, $\omega^\ast$ can be defined by $\omega^\ast\colon F \mapsto F(\mathbb{C})$ for $F \in \EC^\ast$, where $\mathbb{C}$ is the tensor unit of $\vect$. By the proof of Theorem 5 in \cite{ostrik}, it is easy to conclude that $H_{(\EC^\ast, \omega^\ast)} \simeq H_{(\EC, \omega)}^\ast$ as Hopf algebras.  In other words, we have
\begin{equation*}
[\mathrm{ESN}_{(\EC, \omega)}]_\Lambda = [\mathrm{D}(H_{(\EC, \omega)})]_\Lambda \xleftrightarrow{\mbox{EM duality}} 
[\mathrm{D}(H_{(\EC, \omega)}^\ast)]_{\Lambda^\ast} = [\mathrm{ESN}_{(\EC^\ast, \omega^\ast)}]_{\Lambda^\ast}.
\end{equation*}
In tensor-categorical language, the EM duality is a duality between $(\EC, \omega)$ and $(\EC^\ast, \omega^\ast)$. Indeed, it is easy to show\footnote{$\EC^{\ast\ast}=\EC$ follows from Theorem 5 in \cite{ostrik} immediately and $\omega^{\ast\ast}=\omega$ can be easily checked.} that $(\EC^{\ast\ast}, \omega^{\ast\ast}) = (\EC, \omega)$. This categorical duality is related to the Morita equivalence between $\EC$ and its dual $\EC^\ast$. We will explain this point in Section~\ref{sec:defect} where the EM duality is studied from the perspective of domain wall.

\section{From extended string-net models to string-net models} 
\label{sec:esn-to-sn}

We now make the connection with string-net models. String-net models were introduced by Levin and Wen in \cite{LevinWen} and further developed and enriched by boundaries and defects of all codimensions in \cite{KK}. 

The local Hilbert spaces of SN models, solely determined by the data of a unitary tensor category $\EC$, are different from those of ESN models in general. However, the ground levels of SN models and their extended versions are identical. They describe the same topological phases. Given an $[\mathrm{ESN}_{(\EC, \omega)}]_\Lambda$ model (or equivalently a $[\mathrm{D}(H_{(\EC, \omega)})]_\Lambda$ model), one can project out the additional degrees of freedom controlled by the fiber functor $\omega\colon \EC \to \vect$, in a manner entirely analogous to the construction in~\cite{BA}, to obtain a SN model defined by the unitary tensor category~$\EC$. This resulting SN model is called the \emph{magnetic projection} of the original ESN model on $\Lambda$, and will be denoted as $[\mathrm{SN}_\EC^{\mathrm{mag}}]_\Lambda$.

The same construction can be applied to the $[\mathrm{ESN}_{(\EC^\ast, \omega^\ast)}]_{\Lambda^\ast}$ model (or equivalently the $[ \mathrm{D} ( H_{(\EC, \omega)}^\ast )]_{ \Lambda^\ast }$ model). Its magnetic projection is a
string-net model $[ \mathrm{SN}_{\EC^\ast}^{\mathrm{mag}} ]_{ \Lambda^\ast }$. We will call $[ \mathrm{SN}_{\EC^\ast}^{\mathrm{mag}} ]_{ \Lambda^\ast }$ the \emph{electric projection} of $[\mathrm{ESN}_{(\EC, \omega)}]_\Lambda$, 
i.e. $[\mathrm{SN}_\EC^{\mathrm{el}}]_\Lambda \coloneqq [ \mathrm{SN}_{\EC^\ast}^{\mathrm{mag}} ]_{ \Lambda^\ast }$. 

\begin{figure}
\begin{center}
  \includegraphics[scale=1.3]{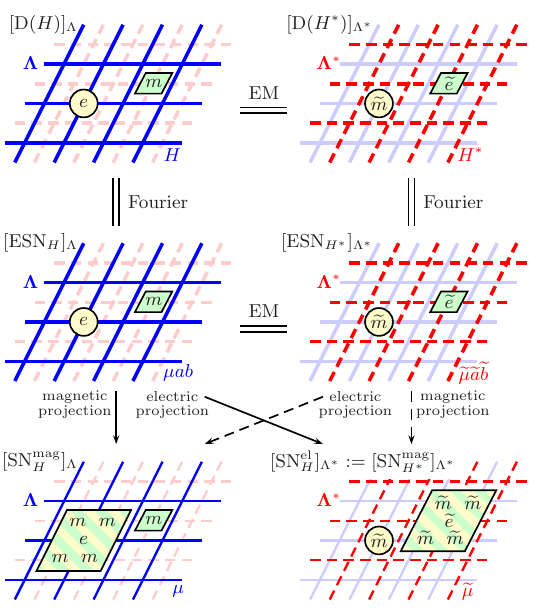}
\end{center}
\caption{\textsl{From quantum doubles to string-nets.}  Quantum double
  models and string-net models are interconnected by EM duality, the
  Fourier mapping, and a projection to SN degrees of freedom as
  illustrated herein and discussed in detail in the text.}
\label{fig:schememodels} 
\end{figure}

The EM duality between $[\mathrm{ESN}_{(\EC, \omega)}]_\Lambda$ and $[\mathrm{ESN}_{(\EC^\ast, \omega^\ast)}]_{\Lambda^\ast}$ induces a duality between $[\mathrm{SN}_\EC^{\mathrm{mag}}]_\Lambda$ and $[ \mathrm{SN}_{\EC^\ast}^{\mathrm{mag}} ]_{ \Lambda^\ast }$, which describe the same topological phase. This net of correspondences and dualities is illustrated in Figure~\ref{fig:schememodels}. 
In this figure, the direct lattice is
drawn solid, the dual lattice dashed.  The first layer represents a
quantum double model in its two EM-dual formulations on the direct and
dual lattices $\Lambda$ and~$\Lambda^\ast$.  Via a Fourier mapping we
write the quantum doubles as extended string-net models, depicted in
the second layer, which are entirely equivalent.  The degrees of
freedom of the ESN model on $\Lambda$ are labelled by irreducible
representations $\mu$ of $H$ and auxiliary matrix indices $a$, $b$;
for the dual model on $\Lambda^\ast$ the labels are irreducible
representations $\widetilde{\mu}$ of $H^\ast$, and auxiliary matrix
indices $\widetilde{a}$, $\widetilde{b}$.  These models support
excitations both at vertices (electric) and at plaquettes (magnetic),
which we represent by blobs and lozenges.  An electric excitation $e$
becomes a magnetic excitation $\widetilde{m}$ in the dual, and
vice versa.  The third layer in Figure~\ref{fig:schememodels}
represents the string-net models obtained, respectively, by projecting
to degrees of freedom $\mu$ on the direct lattice, or to
$\widetilde{\mu}$ on the dual lattice; these we refer to as magnetic
and electric projections of the ESNs, respectively.  Excitations in SN
models are less symmetric than in the ESN: for instance, violation of
a vertex constraint implies the violation of all surrounding plaquette
constraints (see~\cite{BA}). This is due to the fact that the local Hilbert spaces of a SN model is too small to support a well-defined splitting of electric and magnetic excitations.  

For instance, in~\cite{BA} the magnetic SN projection of the $\mathrm{D} (G)$ model was analysed.  Its \emph{electric} SN
projection, on the other hand, has the same local degrees of freedom as the $\mathrm{D} (G)$ model, since group elements are the (one-dimensional) irreps of the dual of the group algebra. 
This connection of SN and $\mathrm{D} (G)$ models had been recognised long ago by H\'ector Bomb\'{i}n \cite{bombin}.

%%%%%%%%%%%%%%%%%%%%%%%%%%%%%%%%%%%%%%%%%%%%%%%%%%%%%%%%%
\section{EM duality and domain walls}
\label{sec:defect} 

\begin{figure}
 \raisebox{-160pt}{
  \begin{picture}(160,160)
   \put(120,8){\scalebox{1.5}{\includegraphics{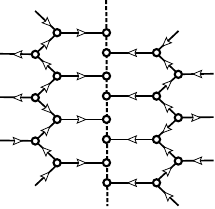}}}
   \put(120,8){
     \setlength{\unitlength}{.75pt}\put(-151,-235){
     \put(228,410)  {\scriptsize $ \mu_1 $}
     \put(228,370)  {\scriptsize $ \mu_2 $}
     \put(228,327)  {\scriptsize $ \mu_3 $}
     \put(228,285)  {\scriptsize $ \mu_4 $}
     \put(240,440)  {\scriptsize $\vect$ }
     %\put(260,420)  {\scriptsize  $\lambda_1 $}
     %\put(260,392)  {\scriptsize $\lambda_2 $}
     %\put(260,372)  {\scriptsize $ \lambda_3 $}
      %\put(260,350)  {\scriptsize $ \lambda_4 $}
     %\put(260,328)  {\scriptsize $ \lambda_5 $}
     %\put(260,308)  {\scriptsize $ \lambda_6$}
     %\put(260,286)  {\scriptsize $ \lambda_7$}
      %\put(260,266)  {\scriptsize $ \lambda_8 $}
       %\put(260,242)  {\scriptsize $ \lambda_9 $}
     \put(278,390)  {\scriptsize $ \nu_1' $}
     \put(278,350)  {\scriptsize $ \nu_2' $}
     \put(278,307)  {\scriptsize $ \nu_3' $}
     \put(278,265)  {\scriptsize $ \nu_4' $}
     }\setlength{\unitlength}{1pt}}
  \end{picture}}
\caption{\label{lw-mod-defect}
A neighborhood of a defect line between a $\EC$-lattice and a $\EC^\ast$-lattice, where $\mu_i\in \EC$ and $\nu_i'\in \EC^\ast$. There is no edge label on the domain wall because there is only one simple object $\mathbb{C}$ in $\vect$.}
\end{figure}
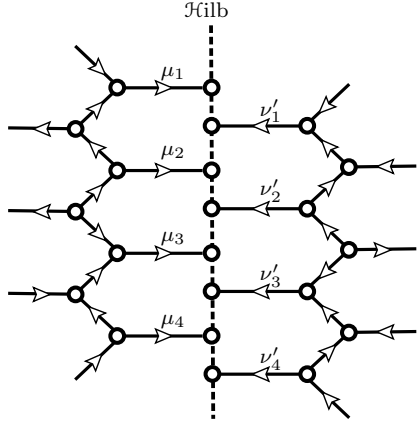

A duality-defect correspondence was obtained in \cite{KK}\footnote{The corresponding mathematical result was first proved independently in \cite{eno}.}. In particular, it was shown that invertible dualities between two bulk SN models based on the unitary tensor categories $\EC$ and $\ED$ are in one-to-one correspondence with the invertible domain walls between these two SN models. 
Therefore, the EM duality between $[\mathrm{SN}_\EC^{\mathrm{mag}}]_\Lambda$ and $[ \mathrm{SN}_{\EC^\ast}^{\mathrm{mag}} ]_{ \Lambda^\ast }$ defined (for a given fiber functor) in Section~\ref{sec:esn-to-sn} must correspond to an invertible domain wall. This domain wall, depicted in Fig. \ref{lw-mod-defect}, is nothing but the category $\vect$, which is viewed as an invertible $\EC$-$\EC^\ast$-bimodule. Moreover, the category $\vect$ defines a Morita equivalence between $\EC$ and $\EC^\ast$, and provides a braided equivalence (defined below by Eq.~\eqref{eq:duality-M} when $\EM=\vect$) from the bulk excitations $Z(\EC)$ to the bulk excitations $Z(\EC^\ast)$. This braided equivalence describes exactly how a bulk excitation in a $\EC$-lattice tunnels through the domain wall into a $\EC^\ast$-lattice \cite{KK}.   

Although we can define the EM duality between two SN models obtained from the magnetic and electric projections of a given ESN model, there is no way to define EM duality for general SN models because (i)~a fiber functor $\EC \to \vect$ does not exist in general (and so a parent ESN model as introduced above does not exist, either), and (ii)~even if such a fiber functor does exist it is not unique in general. In other words, each fiber functor is as canonical as any other and does not deserve to be called \enquote{the} EM duality.

It is, however, possible to associate to every SN model based on a unitary tensor category $\EC$ a family of  dualities parametrized by indecomposable semisimple $\EC$-modules. Indeed, given such a $\EC$-module $\EM$, we can define the dual tensor category $\EC_\EM^\ast\coloneqq \mathrm{Fun}_\EC(\EM, \EM)^{\otimes_{\mathrm{op}}}$ by the category of $\EC$-module functors from $\EM$ to $\EM$ with tensor product defined by the opposite composition of functors. Then there is a duality between the SN model based on $\EC$ and that based on $\EC_\EM^\ast$. This duality~$\alpha_\EM$ is precisely determined by the domain wall (or the defect line) associated to $\EM$ via the duality-defect correspondence \cite{KK}. More precisely, the duality~$\alpha_\EM$, as an invertible braided monoidal functor $\alpha_\EM\colon Z(\EC) \to Z(\EC_\EM^\ast)$ from the bulk excitations~$Z(\EC)$ of a $\EC$-lattice to the bulk excitations $Z(\EC_\EM^\ast)$ of a $\EC_\EM^\ast$-lattice, is defined by: 
\begin{align}
Z(\EC)=\mathrm{Fun}_{\EC|\EC}(\EC, \EC) &\to \mathrm{Fun}_{\EC|\EC}(\EM^{\mathrm{op}}\boxtimes_\EC \EC \boxtimes_\EC \EM, \EM^{\mathrm{op}}\boxtimes_\EC \EC \boxtimes_\EC \EM) \simeq Z(\EC_\EM^\ast) \nonumber \\
\EF &\mapsto \id_{\EM^{\mathrm{op}}} \boxtimes_\EC \EF \boxtimes_\EC \id_{\EM}.  \label{eq:duality-M}
\end{align}
If $\EC$ is equipped with a fiber functor $\omega\colon \EC \to \vect$, then the category $\vect$ is automatically a $\EC$-module. Then the functor $\alpha_\EM$ for $\EM=\vect$ defines the EM duality between the bulk excitations in $[\mathrm{SN}_\EC^{\mathrm{mag}}]_\Lambda$ and those in $[ \mathrm{SN}_{\EC^\ast}^{\mathrm{mag}} ]_{ \Lambda^\ast }$.

\medskip
We have shown that EM duality can be realized by an invertible domain wall (or a defect line) at the level of SN models. We would like to mention that it is also possible to associate EM duality to an invertible domain wall at the level of ESN models. The theory of ESN models enriched by boundaries and defects demands many new mathematical notions and tools, and will be developed elsewhere~\cite{kong2}. We will provide a quick glance here into this rich theory so that we will be able to say something about the EM duality. 

The boundary-defect theory of ESN models relies on a generalization of the representation theory of a tensor category $\EC$ to that of a pair $(\EC, \omega_\EC)$ where $\omega_\EC$ is a fiber functor. In particular, a $(\EC, \omega_\EC)$-module is a pair $(\EM, \omega_\EM)$ where $\EM$ is a $\EC$-module and $\omega_\EM$ is such that the following diagram: 
\begin{equation*}
\xymatrix{
\EC \times \EM \ar[r]^\otimes \ar[d]_{\omega_\EC \times \omega_\EM} & \EM \ar[d]^{\omega_\EM}  \\
\vect \times \vect \ar[r]^\otimes & \vect
}
\end{equation*}
commutes (up to a coherent isomorphism). Such a pair $(\EM, \omega_\EM)$ can be easily constructed from a special symmetric Frobenius algebra $A$ in $\EC$, by setting $\EM$ to be the category $\mathrm{Mod}_A(\EC)$ of $A$-modules internal to $\EC$ and $\omega_\EM$ to be the composition of the functors $\mathrm{Mod}_A(\EC) \hookrightarrow \EC \xrightarrow{\omega_\EC} \vect$. An ESN model with a boundary can be defined by the data $(\EC, \omega_\EC)$ in the bulk and by $(\EM, \omega_\EM)$ on the boundary. Similarly, an ESN model containing a domain wall between a $(\EC, \omega_\EC)$-lattice and a $(\ED, \omega_\ED)$-lattice can be constructed by a pair $(\EM, \omega_\EM)$ where $\EM$ is a $\EC$-$\ED$-bimodule equipped with a fiber functor $\omega_\EM$ compatible with $\EC$-$\ED$-actions, $\omega_\EC$ and $\omega_\ED$. With all of this in mind, the EM duality is then given by the domain wall associated to the pair $(\vect, \id_{\vect})$ where the $\EC$-$\EC^\ast$-bimodule structure on $\vect$ is induced from the monoidal functor $\omega_\EC$. Intuitively, it is just an empty wall in the lattice without any physical degrees of freedom on it, i.e. $\mathcal{H}_e$ is one-dimensional for each internal edge $e$ on the wall. So it shifts a lattice to its dual lattice. This $(\vect, \id_{\vect})$-wall is clearly invertible with the inverse given by itself. 

An example of such a domain wall in the toric code is explicitly given as the vertical dotted line in Fig.~1 in \cite{KK}. In that case, by applying local operators given by $\sigma$-matrices, one can see explicitly how an electric charge tunnels through the wall and becomes a magnetic flux. This tunneling process gives the EM duality in the toric code model. 

%%%%%%%%%%%%%%%%%%%%%%%%%%%%%%%%%%%%%%%%%%%%%%%%%%%%%%%%%
\section{Measuring topology with tensor networks}
\label{sec:phys} 

Some physical consequences of EM duality are immediate.  First of all,
EM duality allows us to measure the topology of the surface underlying
the lattice models by using only locally defined states.

As shown in~\cite{BMCA}, each $[ \mathrm{D} (H) ]_\Lambda$ model has one
canonical tensor network ground state $| \psi ( H, \, \Lambda )
\rangle$ constructed from \emph{identical} tensors at each site, which
are defined solely by the structure of $H$.

{}From the corresponding canonical state $| \psi ( H^\ast, \,
\Lambda^\ast ) \rangle$ of the dual $[ \mathrm{D} (H^\ast) ]_{
  \Lambda^\ast }$ model on the dual lattice we can obtain another
ground state of the original $[ \mathrm{D} (H) ]_\Lambda$ model by
using the duality map~\eqref{emduality}:
\begin{equation}\label{dualgroundstate}
  | \widetilde{\psi} ( H, \, \Lambda ) \rangle
=
  U_\Lambda^\dagger \, 
  | \psi ( H^\ast, \, \Lambda^\ast ) \rangle
\; .
\end{equation}

The relation between $| \psi ( H, \, \Lambda ) \rangle$ and
$| \widetilde{\psi} ( H, \, \Lambda ) \rangle$ depends on the
topology of the surface underlying $\Lambda$.  On the sphere, for
instance, the ground level is nondegenerate, so both ground states are
the same.  On a topologically nontrivial surface such as the torus,
these ground states can be shown to always be linearly independent.

For instance, in the toric code the tensor network construction of the
canonical state $| \psi ( H, \, \Lambda) \rangle$ coincides with the
\emph{projected entangled-pair state} (PEPS) ansatz
in~\cite{VerstraetePower}.  On the torus this yields the logical $| ++
\rangle$ state; the dual ground state $| \widetilde{\psi} ( H, \,
\Lambda ) \rangle$ is the logical $| 00 \rangle$ state.

On the other hand, the correspondence between string-net and quantum
double models can be used to relate these PEPS to the tensor network
descriptions of string-net ground states put forward in~\cite{BAV}
and~\cite{GuTopo}.  These come from the $| \psi ( H, \, \Lambda )
\rangle$ construction of the corresponding $[ \mathrm{D} (H)
]_\Lambda$ model, while the $| \widetilde{\psi} ( H, \, \Lambda )
\rangle$ can be seen to yield a \emph{new} TN given simply by the
knitting together of $3j$ symbols at vertices in the ESN degrees of
freedom.  The Fourier construction is also expected to relate to the
\emph{entanglement renormalization} analyses of $\mathrm{D}
(H)$~\cite{AguadoVidal, BMCA} and SN models~\cite{Koenig}.

%%%%%%%%%%%%%%%%%%%%%%%%%%%%%%%%%%%%%%%%%%%%%%%%%%%%%%%%%
\section{Discussion and outlook}
\label{sec:discussion} 

We have defined the EM duality for non-Abelian $\mathrm{D} (G)$ models,
showing how it arises naturally in the context of $\mathrm{D} (H)$
models.  The connection to SN models comes from the Fourier
construction of ESN models and their electric and magnetic
projections. We have also shown that the EM-duality can be realized by an invertible domain wall. 
The EM duality offers nonlocal information about the systems, unveiling
global characteristics of space, from tensor networks defined locally.
Beyond this, it serves as an \emph{organising principle} for
topological models; indeed, we envisage a net of dualities for
topological phases (not unlike that connecting theories of strings and
branes).  This then reflects on the field theories underlying
topological systems and on their mathematical structures, such as
tensor category theory.

We wish to stress the importance of the whole class of ESN models.  This is indeed the natural habitat to study electric-magnetic duality among lattice models for intrinsic, non-chiral topological orders.  
We thus shift the focus away from lattice gauge theories based on groups, which are a primary concern in high energy theory, even when the Hopf algebra language is applied~\cite{Oeckl}.  
Ours is, moreover, the first proposal for a general EM duality governing topologically ordered systems.

\medskip
Nevertheless there is a remaining issue with the magnetic projection from ESN models to SN models: it is not onto because not all finite fusion categories have fiber functors or, equivalently, come from the representations of a semisimple Hopf algebra. However, for every unitary finite fusion category $\EC$ there exists a functor $\omega\colon \EC \to \vect$ which is equipped with a separable Frobenius structure \cite{hayashi,pfeiffer}. Such a structure is weaker than a monoidal functor, yet it is good enough to construct projectors $A_v$ and $B_p$ in a way similar to the construction in Section~\ref{sec:qdsn}. Correspondingly, the space $\mathcal{H}_e=\oplus_{\mu\in I} \omega(\mu) \otimes \omega(\mu)^\ast$ has a natural structure of a weak $C^\ast$-Hopf algebra. So in order to recover all SN models via magnetic projection, we need to generalize the construction of quantum double models to weak $C^\ast$-Hopf algebras. We believe that such a generalization is possible and will leave it to future study. 

\medskip
Finally, the nature of the anyonic models controlling the excitations of the $\mathrm{D} (H)$ models (or ESN models) has a bearing on the experimental realisation of string-net models, which we contend would be most natural in their~\emph{extended} incarnations. In particular, protocols for anyonic manipulation in the spirit of~\cite{brennenetal} and~\cite{brennenetaltwo} would then be possible.  Moreover, the invertible domain wall associated to the EM duality is of the simplest kind. There is no physical degree of freedom associated to the edges on the wall. If an ESN model is ever to be realized in a lab, we expect that the EM domain wall will be among the first types of walls to be observed. A further key question to be explored is the meaning of EM duality under perturbation of the models considered, that is, away from topological fixed points; notice that notions of duality different from the EM duality discussed here are relevant for the toric code and its perturbations, as shown in \cite{JVidal}.

\bigskip
\noindent \textbf{Acknowledgements:} We would like to thank H.~Bomb\'{\i}n,
J.~M.~Mombelli, G.~Ortiz, J.~Slingerland, and G.~Vidal for
discussions.
MC acknowledges financial support by the German Science Foundation
(grant CH 843/2-1), the Swiss National Science Foundation (grants
PP00P2\_128455, 20CH21\_138799 (CHIST-ERA project CQC)), the Swiss
National Center of Competence in Research `Quantum Science and
Technology (QSIT)' and the Swiss State Secretariat for Education and
Research supporting COST action MP1006.
LK
is supported by the Gordon and Betty Moore Foundation through CPI in
Caltech, NSF under Grant No.~PHY-0803371, the Basic Research
Young Scholars Programs, the Initiative Scientific Research Program of Tsinghua University, 
and NSFC under Grant No. 11071134.
Research at Perimeter Institute is supported by the Government of
Canada through Industry Canada and by the Province of Ontario through
the Ministry of Research.

%%%%%%%%%%%%%%%%%%%%%%%%%%%%%%%%%%%%%%%%%%%%%%%%%%%%%%%
%%%%%%%%%%%%%%%%%%%%%%%%%%%%%%%%%%%%%%%%%%%%%%%%%%%%%%%

\end{document}